\documentclass[prl,twocolumn,superscriptaddress,floatfix,showpacs]{revtex4}

\usepackage{amssymb,amsmath,amsthm,graphicx,ulem}
\pdfoutput=1
\usepackage{graphicx,subfigure}
\usepackage{ulem}
\usepackage{epsfig}
\usepackage{amsmath}
\usepackage{amsfonts}
\usepackage{amssymb}
\usepackage[usenames]{color}
\usepackage[letterpaper,left=2.cm,right=2.cm,top=2.5cm,bottom=2.5cm]{geometry}
\usepackage[T1]{fontenc}

\newcommand{\beq}{\begin{equation}}
\newcommand{\eeq}{\end{equation}}
\newcommand{\be}{\begin{equation}}
\newcommand{\ee}{\end{equation}}
\newcommand{\beqa}{\begin{eqnarray}}
\newcommand{\eeqa}{\end{eqnarray}}
\newcommand{\beqar}{\begin{eqnarray*}}
\newcommand{\eeqar}{\end{eqnarray*}}
\newcommand{\bea}{\begin{eqnarray}}
\newcommand{\eea}{\end{eqnarray}}

\newcommand{\ra}{\rightarrow}

\newcommand{\nn}\nonumber
\newcommand{\eqn}[1]{(\ref{#1})}

\begin{document}

\title{Holographic Signatures of Cosmological Singularities}

\author{Netta Engelhardt}
\affiliation{Department of Physics, UCSB, Santa Barbara, CA 93106, USA}
\author{Thomas Hertog}
\affiliation{Institute for Theoretical Physics, KU Leuven, 3001 Leuven, Belgium}
\author{Gary T. Horowitz}
\affiliation{Department of Physics, UCSB, Santa Barbara, CA 93106, USA}


\bibliographystyle{unsrt}

\begin{abstract}
\noindent 

To gain insight into the quantum nature of cosmological singularities, we study anisotropic Kasner solutions in gauge/gravity duality. The dual description of the bulk evolution towards the singularity involves ${\cal N}=4$ super Yang-Mills on the expanding branch of deformed de Sitter space and is well defined. We compute two-point correlators of Yang-Mills operators of large dimensions using spacelike geodesics anchored on the boundary. The correlators show a strong signature of the singularity around horizon scales and decay at large boundary separation at different rates in different directions. More generally, the boundary evolution exhibits a process of particle creation similar to that in inflation. This leads us to conjecture that information on the quantum nature of cosmological singularities is encoded in long-wavelength features of the boundary wave function.

\end{abstract}

\maketitle 
 



\centerline{\bf Introduction}
\vskip .3cm
\noindent
A longstanding goal of quantum gravity is to describe physics near  singularities like the big bang or inside black holes.  Gauge/gravity duality is a  powerful tool to apply to this problem since it maps it into a problem in ordinary QFT on a fixed spacetime background. 
To model a cosmological singularity using holography, one needs to construct an asymptotically anti-de Sitter (AdS) solution to Einstein's equation that evolves into (or from) a singularity which extends all the way out to infinity.
This was first done in \cite{Hertog:2004rz,Turok:2007ry}, but the dual field theory itself became singular when the bulk singularity hit the boundary. In \cite{Maldacena:2010un,Harlow:2010my} the same singular bulk solutions were reinterpreted as being dual to a well defined field theory on de Sitter (dS) spacetime. However it is not clear how (and indeed whether) the dual field theory on dS describes the region near the singularity. This is because the probes which are best understood, such as extremal surfaces which end on the boundary, do not probe the region near the singularity \cite{Engelhardt:2013tra}. Models of this type were further explored in \cite{Barbon:2011ta} and other models were studied in \cite{Craps:2006xq,Das:2006dz,Awad:2008jf}.

Attempts to probe the black hole singularity were somewhat more successful in that there are geodesics with endpoints on the boundary which get arbitrarily close to the singularity \cite{Fidkowski:2003nf}. Unfortunately, it was shown that the two point correlator is not dominated by these geodesics, although their effects could be seen by analytic continuation \cite{liu}.
Nevertheless the presence of the black hole horizon means that clear signatures of the singularity have remained difficult to identify in the dual. 

The goal of this paper is to introduce a new holographic model of a cosmological singularity which has the advantages that 1) the dual field theory is simply strongly-coupled ${\cal N}=4$ super Yang-Mills with a large number $N$ of colors on an anisotropic generalization of de Sitter and is manifestly well defined for all time, and 2) there are bulk geodesics with endpoints on the boundary which come close to the singularity. As a bonus, one can solve for the equal time correlator analytically. We indeed find distinctive behavior which, we argue, signals the presence of the bulk singularity \cite{sethi}. While singularities are ultimately described by quantum gravity, \textit{i.e.} the small $N$ regime, obtaining a field theory description of a classical (large $N$) singularity is an important pioneering step in recasting the problem of singularities in quantum gravity in terms of the dual field theory. In particular, the transition from large to small $N$ is a tractable problem in the field theory, but remains poorly-understood in the bulk.


\vskip .3cm
{\bf \centerline {The solution}}
\vskip .3cm
\noindent
Solutions to Einstein's equation in five dimensions with negative cosmological constant can be obtained by starting with $ AdS_5$ in Poincare coordinates, and replacing the flat Minkowski metric on each radial slice with any Ricci flat metric. The Kasner metric 
\begin{equation} 
ds^{2}= -dt^{2} + t^{2p_{1}}dx_1^{2} + t^{2p_{2} }dx_{2}^{2} + t^{2p_{3}}dx_{3}^{2} \label{BoundaryMetric}
\end{equation}
with  $\sum\limits_{i} p_{i} = 1 = \sum\limits_{i} p_{i}^{2}$ is a well known Ricci flat metric describing a homogeneous, but anisotropic cosmology. It has a 
 singularity in the Weyl curvature
  at $t=0$. With this metric on each radial slice of $AdS_5$, we obtain \cite{Das:2006dz}
 \begin{equation} 
 ds^{2}= \frac{1}{z^{2}}\left ( -dt^{2} + t^{2p_{1}}dx_1^{2} + t^{2p_{2} }dx_{2}^{2} + t^{2p_{3}}dx_{3}^{2} +dz^2 \right)\label{Kasneton} 
 \end{equation}
where we have set the AdS radius to one. It might appear that the dual would have to live on a Kasner spacetime. 
However we can divide the metric in parenthesis by  $H^2 t^2$ where $H$ is some constant, and replace the overall conformal factor by $H^2 t^2/z^2$.
 Writing $Ht = e^{H\tau}$, $x_i = H^{p_i}y_i$ this yields the boundary metric 
\be\label{anidS}
ds^2 =  -d\tau^2 + \sum_i e^{-2(1-p_i) H\tau} dy_i^2
\ee
which is an anisotropic deformation of dS space in flat slicing. In addition to the obvious translational symmetries, \eqn{Kasneton} is invariant under a dilation symmetry \cite{lifshitz}:
\be 
z \ra  \lambda z, \quad t \ra  \lambda t,  \quad x_i \ra \lambda^{(1-p_i)} x_i
\label{dilation}
\ee
This leaves the conformal factor $Ht/z$ invariant and thus acts as an isometry of the boundary metric \eqn{anidS}.

\vskip .3cm
\centerline{\bf Two-point correlator}
\vskip .3cm
\noindent
\noindent In the large $N$ limit, the leading contribution to the two-point correlator of an operator $\mathcal{O}$ of high conformal dimension $\Delta$ in the dual strongly coupled $SU(N)$ Yang-Mills theory on \eqn{anidS} is given by the (regulated) length of spacelike bulk geodesics connecting the two points:
\begin{equation} \left \langle\psi \right | \mathcal{O}\left(x\right)\mathcal{O}\left(x'\right)\left | \psi \right\rangle =e^{-m \mathcal{L}_{reg}(x,x')}\end{equation}
\noindent where $\left | \psi \right\rangle$ is the state of the Yang-Mills theory, $m$ is the mass of the bulk field that is dual to the boundary operator $\mathcal{O}$, and $\mathcal{L}_{reg}(x,x')$ is the regularized length of the bulk geodesic. When $\mathcal{O}$ is a scalar operator, we have $\Delta = 2 + \sqrt{4 + m^{2}}$.

The length of spacelike geodesics is infinite.  As usual, we regulate this length by introducing a cut-off when the conformal factor becomes large, and subtracting the divergent contribution from pure AdS. Writing $\tilde z = z/Ht$, our cut-off will be $\tilde z = \tilde \epsilon$. 

We can solve for the bulk geodesics using the metric \eqn{Kasneton}. We consider equal-time correlators for two points separated in the $x_1$ direction only (hereafter referred to as $x$). The dilation symmetry \eqn{dilation} together with the translational symmetry in $x_2$ and $x_3$ imply that the correlators depend only on the proper boundary separation $\mathcal{L}_{bdy}$ between the two points, and of course on the exponent $p_1$ which we hereafter denote as $p$. Without loss of generality we take the endpoints at $z=0$ to be $\{t=1, \ x=\pm \bar x\}$.
Using $t$ as a parameter, the geodesic equations are: 
\begin{align}
x''(t)t= &p x'(t)\left[-2 + t^{2 p} x'(t)^{2}\right] \label{eoms1}  \\
z''(t)z(t)= &1 -  z'(t)^{2} - t^{2 p-1} x'(t)^{2} \left[t - p z(t)z'(t)\right]
\label{eoms2}
\end{align}
The solutions of \eqn{eoms1} are hypergeometric functions for all $p$.
For $p = \pm 1/n$, with integer $n$, the hypergeometric functions simplify, which makes the analysis more tractable.
We first compute the correlator in a simple nonsingular example, before treating the case $p=-1/4$ that describes an anisotropic dS boundary dual to a bulk with a genuine curvature singularity.

\vskip .2cm
\centerline{\it Correlators in the Milne Universe}
\vskip .2cm
\noindent
The Milne solution is a special case of the Kasner solution \eqn{Kasneton} where one of the $p_{i}=1$ and the rest are zero. This metric features a coordinate singularity at $t=0$, and is simply flat space in alternative coordinates. If $p=0$, the effective $2+1$-dimensional metric determining geodesic motion is precisely AdS$_{3}$. Hence with our choice of boundary conditions the geodesics lie entirely in the surface $t=1$. In terms of the usual cut-off $z=\epsilon$, their length is
$\mathcal{L} = 2 \ln \left ({2\bar x}/{\epsilon} \right) = 2 \ln(\mathcal{L}_{bdy}/\epsilon)$ where $\mathcal{L}_{bdy}$ is the proper boundary separation on the Minkowski boundary. With a cut-off $ \tilde{\epsilon} =  \epsilon/H$ appropriate for 
a boundary de Sitter metric
\begin{equation}
\mathcal{L} = 2 \ln \left (\frac{2\bar x}{H} \cdot \frac{H}{{\epsilon}} \right)=2 \ln \left ( \mathcal{L}_{bdy}\right) -2 \ln \left (\tilde{\epsilon}\right) 
\end{equation}
where $\mathcal{L}_{bdy}$ is now the proper boundary separation on the de Sitter boundary. 
Hence the correlator for a large dimension operator in a $p=0$ direction is given by
\begin{equation} \label{standard}
\left \langle \mathcal{O}(\bar x)\mathcal{O}(-\bar x)\right\rangle_{p=0} = \mathcal{L}_{bdy}^{-2\Delta}.\end{equation}
Note that the result is the same as flat space and independent of $H$, as expected for a conformal field theory on a conformally flat spacetime.

For $p=1$ the effective 2+1 metric seen by a geodesic 
can be transformed into pure AdS$_3$ in Poincare coordinates by the coordinate transformation
$(t,x) \rightarrow (\eta,\chi) = (t \cosh x,t \sinh x)$.
Using this, we can obtain the length of a geodesic anchored at $x=\pm \bar x$ and $t=1$ from the result for $p=0$. This yields the following 
equal time correlator 
\be
\left \langle \mathcal{O}(\bar x)\mathcal{O}(-\bar x)\right\rangle_{p=1} =\left[{2\over H}\sinh\left(\frac{H}{2} \mathcal{L}_{bdy}\right)\right]^{-2\Delta}
\ee
which falls off exponentially with proper distance. 
This is precisely the correlator in a thermal state with temperature $T = H/2\pi$.

\vskip .2cm
\centerline{\it Correlators in Anisotropic de Sitter}
\vskip .2cm
\noindent
We now turn to our central example $p=-1/4$ which describes a genuinely singular bulk solution. We will set $H=1$ for convenience. For $p=-1/4$ the solutions of \eqn{eoms1}--\eqn{eoms2} can be written as
\begin{align}
&x(w) =  \frac{4}{15}\sqrt{c + w} (8 c^{2} - 4 c w + 3 w^{2})
\label{soln1/4}\\
&z(w) = \frac{4}{3} \sqrt{c [ w^3 -1 + 3 c (1 - w^{2})]}
\label{soln1/4b}
\end{align}
where $w=\sqrt{t}$ and $c$ is an integration constant. The solutions \eqn{soln1/4}--\eqn{soln1/4b} describe half of the geodesics from the boundary at $w=1$ and $x=\bar x$ up to a turning point in the interior at $w=w_{*}$ where $x=0$.
At the turning point $dt/dx = 2wdw/dx =0$ which implies $dx/dw \rightarrow \infty$, so $w_{*} = -c$. 

\begin{figure}[t]
\includegraphics[width=3in]{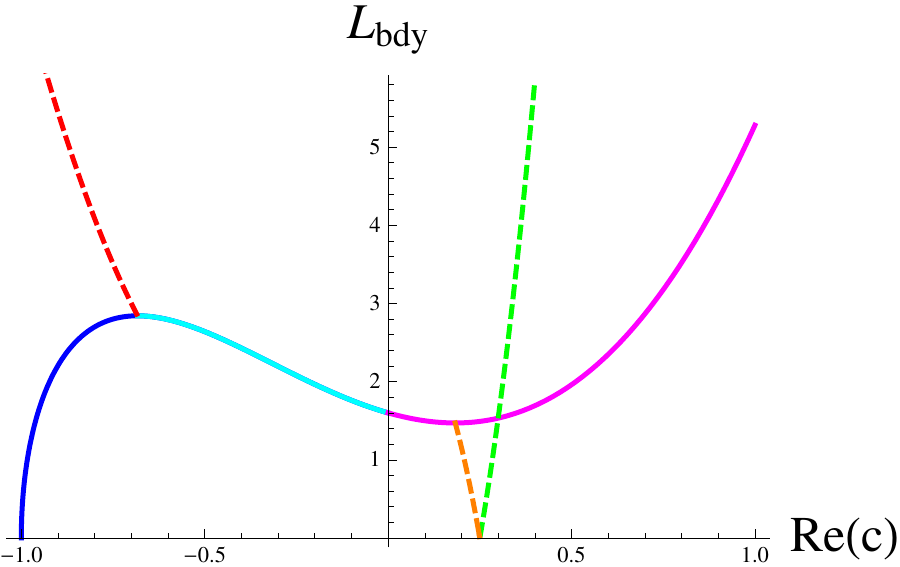} 
\caption{The proper boundary separation $\mathcal{L}_{bdy}$ as a function of the real part of $c$ for $p=-1/4$. The solid curve corresponds to real $c$ whereas the dashed  curves correspond to complex conjugate pairs of $c$. Note that there are five possible geodesics for each $\mathcal{L}_{bdy}$.}
\label{lbdy}
\end{figure}
Since $\mathcal{L}_{bdy} =2 x(1)$ is quintic in $\sqrt c$ there are five, possibly complex geodesics \cite{bala} corresponding to each boundary separation $\mathcal{L}_{bdy}$, which we require to be real and positive. We must determine which ones contribute to the correlator. Fig. \ref{lbdy} shows $\mathcal{L}_{bdy}$ as a function of the real part of $c$. When $\text{Re}(c)>-1$, the geodesics curve towards the singularity, and $\text{Re} (c) >0$ geodesics even propagate all the way through $t=0$ before turning around \cite{sing}. However we must discard the contributions from $\text{Re}(c)> 0$ geodesics because they would predict that the correlator increases as the separation between the two points grows, and they would result in an unphysical pole on a spacelike surface on the boundary. Moreover, the geodesic approximation is  only justified where the spacetime is analytic \cite{Louko:2000tp}, and our solution is certainly not analytic at $t=0$. 
The net result is that the real geodesics of interest have $-1<c<0$ and $-c \le w \le 1$. As $c\to 0$ the geodesics approach the singularity.

The length of the geodesic is given by the following contour integral in the complex $w$ plane 
\begin{equation}  \int dw \frac{3 \sqrt{1-3c+4c^{3}} w}{\sqrt{(c+w)}(1-w)[ w^{2} +(1-3c)(w+1)]}
\label{lw}
\end{equation}
from $w=-c$ to $w=\sqrt{1-\delta}$, where $\delta$ is given by the UV cut-off $\tilde{\epsilon} = z(1-\delta)$. (Since $H = t = 1$, our dS cut-off agrees with the standard cut-off in $z$.)
The integral \eqn{lw} has four singularities, at $w=1$, $w=-c$, and two simple poles  at $w_\pm=\frac{1}{2}\left (3c-1 \pm \sqrt{3(3c-1)(c+1)}\right)$. For $c$ real and negative we may directly integrate \eqn{lw} along the real axis, since the poles at $w_\pm$ do not lie on the contour of integration. When $c$ is complex, one simply deforms the contour into the complex plane.
Restricting to Re$(c)<0$ the integral gives
\be
\mathcal{L} = 2 \tanh^{-1} \left [\frac{\left(2c-\sqrt{1-\delta} \right)\sqrt{c+\sqrt{1-\delta}}
}{\sqrt{1+c} (2c-1)} \right]
\ee
which results in the following regulated length
\be \mathcal{L}_{reg} = \ln \left [ - \frac{64}{9} c(1+c)(2c-1)^{2}\right] 
\label{length}.
\ee
The divergence of $\mathcal{L}_{reg}$ at $c=-1 $ is easily seen to be the usual short distance singularity of the correlator: $\mathcal{L}_{bdy} = 8\sqrt{  1+c}$ for small $\mathcal{L}_{bdy}  $, so $\mathcal{L}_{reg}=  2\ln \mathcal{L}_{bdy}$.

\begin{figure}[h]
\includegraphics[width=1.4in]{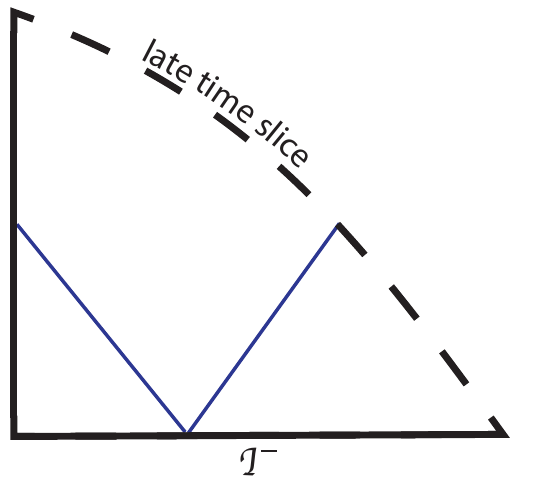}
\caption{A conformal diagram of the anisotropic de Sitter boundary geometry shows that two points separated by the horizon size can be connected by a null geodesic that ``bounces" off $\cal I^{-}$.}
\label{bdy}
\end{figure}

Now consider  the divergence at $c=0$. This  occurs when the boundary separation reaches the cosmological horizon size $\mathcal{L}_{hor}$. For $\mathcal{L}_{bdy}$ slightly larger than $\mathcal{L}_{hor}$, there are bulk geodesics which come close to the singularity before returning to the boundary. As 
$\mathcal{L}_{bdy} \rightarrow \mathcal{L}_{hor}$, these geodesics approach a null geodesic lying entirely in the boundary which ``bounces" off ${\cal I}^-$
(see Fig. \ref{bdy}).

In \cite{Fidkowski:2003nf}, a pole in the correlator  was found corresponding to a geodesic that bounces off a black hole singularity. It was argued that this did not dominate the correlator and could only be seen by analytic continuation to a second Riemann sheet. In contrast, we believe that the pole we see at the horizon scale is physical. This is because 1) we are not in a thermal state, so there is no general argument that such a pole cannot occur, and 2) our divergence is associated with a null geodesic in the boundary and not the bulk. Physically, the pole at the horizon scale indicates that the initial state of the field theory, which describes the bulk singularity, contains particles created at each point on ${\cal I}^-$, moving in opposite directions. At all later times, these particles will be separated by the horizon scale.

The pole at the horizon scale in the correlator is $(\mathcal{L}_{bdy} - \mathcal{L}_{hor})^{-\Delta}$ which is weaker than the pole at short distances which is $\mathcal{L}_{bdy}^{-2\Delta}$. This is consistent with general properties of quantum field theory.

\begin{figure}[t]
\includegraphics[width=3in]{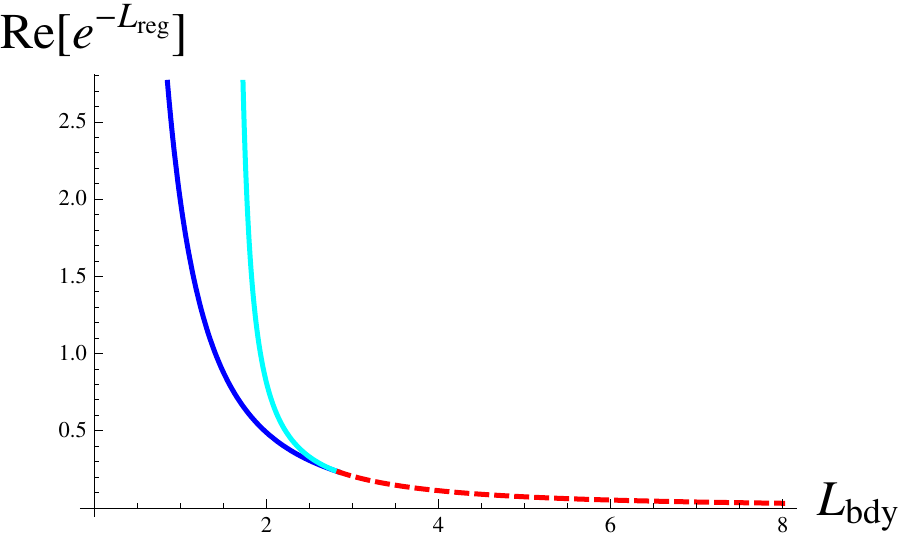}
\caption{Re$\left(e^{-\mathcal{L}_{reg}}\right)$ as a function of the boundary separation for $p=-1/4$, computed from the bulk spacelike geodesics with Re$(c)<0$. The colors correspond to those in Fig. 1, e.g.,  the red dashed line is the contribution from each of the two complex conjugate  geodesics.}
\label{correlator}
\end{figure}

The contributions to the equal time correlator from the one or two geodesics with Re$(c)<0$ are shown in Fig. \ref{correlator}.
 At small boundary separation we obtain the requisite divergence of $\mathcal{L}_{bdy}^{-2\Delta}$ from one real geodesic. At the horizon size, a second real geodesic appears  and produces the pole. At approximately twice the horizon size, the two real geodesics merge and are replaced by complex 
  conjugate geodesics.  As $\mathcal{L}_{bdy}\rightarrow \infty$, its dependence on $c$ simplifies to $\mathcal{L}_{bdy}\propto c^{5/2}$, and we find that the asymptotic two-point correlator has a different fall-off from the correlator in pure de Sitter:
\begin{equation} \left\langle \mathcal{O}(\bar x)\mathcal{O}(-\bar x)\right\rangle_{p=-1/4}\propto \mathcal{L}_{bdy}^{- {8 \Delta}/{5}}.\end{equation}
\noindent Correlations in the $x$ direction 
are therefore \textit{enhanced} in the large separation limit in comparison with correlations in de Sitter. This difference is clearly due to the anisotropy, and by extension, the bulk singularity. 

The behavior for $p = -1/4$ is typical of $p < 0 $. One can show \cite{us} that when $p < 0 $, geodesics always bend toward the singularity, and there always exists a family of spacelike geodesics which turn around close to the singularity. As $\mathcal{L}_{bdy} \rightarrow \mathcal{L}_{hor}$ these geodesics approach a null geodesic lying entirely in the boundary. For $p > 0$, the bulk geodesics bend away from the singularity, so they do not approach a null geodesic on the boundary at the horizon size. As a result, the correlator does not have a pole at the horizon scale in this case.

 For general $p < 1$, the power law falloff with large boundary separation appears to satisfy:
\be \left \langle \mathcal{O}\left ( \bar x\right) \mathcal{O}\left(-\bar x\right)\right\rangle \propto \mathcal{L}_{bdy}^{-{2\Delta}/(1-p)}\label{genp}
\ee
 This holds in all cases we have checked, but we do not yet have a general derivation.
A suggestive way to view this is the following: Our dilation symmetry implies that the general equal time correlator $\left \langle \mathcal{O}\left ( \bar x, \bar t\right) \mathcal{O}\left(-\bar x, \bar t\right)\right\rangle$ is only a function of one   variable $\xi = \bar t/\bar x^{1\over 1-p}$ \cite{ross}.
 Eq. \eqn{genp} states that for small $\xi$, this function is simply $\xi^{2\Delta}$.
 We emphasize that this is {\it different} from the short distance behavior which is always given by \eqn{standard}.

\vskip .3cm
\centerline{\bf Discussion}
\vskip .3cm
\noindent
We have put ${\cal N}=4$ super Yang-Mills on an anisotropic deformation of de Sitter space, and studied the two-point function of a high dimension operator in a state dual to a cosmological singularity in the bulk. We have found two unusual features: in directions with $p<0$, there is a pole precisely at the horizon scale; and the large distance fall-off is a power law with a power that depends on the local expansion rate.  Further details and explorations of the bulk cosmological singularity using different holographic probes will be given elsewhere \cite{us}. Since the inside region of black holes is like an anisotropic cosmology, our setup may also be useful to better understand black hole singularities.

We have focussed on the singularity at $t=0$, but our model contains 
another more subtle singularity 
at the Poincare horizon, $z=\infty$. This can be viewed as a (null) ``big crunch'' singularity in the future. Alternatively, it can be removed by adding one compact dimension and starting with a six dimensional AdS soliton. One can again replace the Minkowski slices with Kasner and have a big crunch in the bulk, however now the bulk smoothly ends at finite $z$ \cite{Engelhardt:2013jda}. Our results about the pole will not be affected since they only depend on geodesics near the boundary, but the large distance fall-off will certainly be modified since one now is in a confining vacuum.


So far we have discussed solutions with an initial `big bang' singularity. However our results also apply to Kasner-AdS solutions with a singularity in the future. The bulk evolution from regular initial data towards the future singularity will then have a dual description in terms of ${\cal N}=4$ super Yang-Mills on a deformed dS space expanding at different rates in different directions.

The anisotropic expansion of the boundary background breaks conformal invariance and gives rise to particle creation, just like a rolling scalar does in inflation in cosmology \cite{skenderis}. The relevant length scales in this process are the expansion rates in different directions. By analogy with inflation one expects that fluctuations will be in their ground state on scales below these, but exhibit particle-like excitations on larger scales. This expectation is born out by the form of the two-point correlator \eqn{genp}. For sub-horizon boundary separations the correlator is at all times close to that in exact dS space. By contrast, on scales larger than the horizon in a given direction it deviates significantly from the correlator in dS reflecting the excited state on those scales.

Hence we are led to a picture in which the holographic dual of cosmological singularities is given in terms of a boundary wave function that describes an ensemble of highly excited configurations on horizon and super horizon scales in an anisotropic de Sitter space. It thus appears that signatures of the quantum nature of cosmological singularities can be found in the classical long-wavelength features predicted by the boundary theory.
\vskip .3cm

\noindent{\bf Acknowledgements:} It is a pleasure to thank D. Berenstein, E. Dzienkowski, S. Fischetti, S. Hollands,  D. Marolf,  S. Ross, and M. Taylor for helpful discussions. This work is supported in part by the US NSF Graduate Research Fellowship under Grant No. DGE-1144085, by NSF Grant No. PHY12-05500 and by the National Science Foundation of Belgium under the FWO-Odysseus program. TH thanks the KITP and the Physics Department at UCSB for their hospitality.


\begin{thebibliography}{99}
  
  
\bibitem{Hertog:2004rz}
  T.~Hertog and G.~T.~Horowitz,
  ``Towards a big crunch dual,''
  JHEP {\bf 0407} (2004) 073
  [hep-th/0406134];
   ``Holographic description of AdS cosmologies,''
  JHEP {\bf 0504}, 005 (2005)
  [hep-th/0503071].
  
\bibitem{Turok:2007ry} 
  N.~Turok, B.~Craps and T.~Hertog,
  ``From big crunch to big bang with AdS/CFT,''
  arXiv:0711.1824 [hep-th];
  ``On the Quantum Resolution of Cosmological Singularities using AdS/CFT,''
  Phys.\ Rev.\ D {\bf 86}, 043513 (2012)
  [arXiv:0712.4180 [hep-th]].



\bibitem{Maldacena:2010un}
  J.~Maldacena,
  ``Vacuum decay into Anti de Sitter space,''
  arXiv:1012.0274 [hep-th].
  
\bibitem{Harlow:2010my}
  D.~Harlow and L.~Susskind,
  ``Crunches, Hats, and a Conjecture,''
  arXiv:1012.5302 [hep-th].

  
\bibitem{Engelhardt:2013tra} 
  N.~Engelhardt and A.~C.~Wall,
  ``Extremal Surface Barriers,''
  JHEP {\bf 1403}, 068 (2014)
  [arXiv:1312.3699 [hep-th]].

\bibitem{Barbon:2011ta}
  J.~L.~F.~Barbon and E.~Rabinovici,
  ``AdS Crunches, CFT Falls And Cosmological Complementarity,''
  JHEP {\bf 1104} (2011) 044
  [arXiv:1102.3015 [hep-th]].
  
\bibitem{Craps:2006xq}
  B.~Craps, A.~Rajaraman and S.~Sethi,
  ``Effective dynamics of the matrix big bang,''
  Phys.\ Rev.\ D {\bf 73} (2006) 106005
  [hep-th/0601062].

\bibitem{Das:2006dz}
  S.~R.~Das, J.~Michelson, K.~Narayan and S.~P.~Trivedi,
  ``Time dependent cosmologies and their duals,''
  Phys.\ Rev.\ D {\bf 74} (2006) 026002
  [hep-th/0602107].
  
\bibitem{Awad:2008jf}
  A.~Awad, S.~R.~Das, S.~Nampuri, K.~Narayan and S.~P.~Trivedi,
  ``Gauge Theories with Time Dependent Couplings and their Cosmological Duals,''
  Phys.\ Rev.\ D {\bf 79} (2009) 046004
  [arXiv:0807.1517 [hep-th]].

\bibitem{Fidkowski:2003nf}
  L.~Fidkowski, V.~Hubeny, M.~Kleban and S.~Shenker,
  ``The Black hole singularity in AdS / CFT,''
  JHEP {\bf 0402} (2004) 014
  [hep-th/0306170].
  
  \bibitem{liu}
  New operators which see the singularity more directly were found in \cite{Festuccia:2005pi}. See also \cite{Hubeny:2013dea} for a recent attempt.
  
\bibitem{Festuccia:2005pi}
  G.~Festuccia and H.~Liu,
  ``Excursions beyond the horizon: Black hole singularities in Yang-Mills theories. I.,''
  JHEP {\bf 0604} (2006) 044
  [hep-th/0506202].
  
    
\bibitem{Hubeny:2013dea}
  V.~E.~Hubeny and H.~Maxfield,
  ``Holographic probes of collapsing black holes,''
  arXiv:1312.6887 [hep-th].
  
  \bibitem{sethi}
Some of this behavior has also been seen in the dual of a null bulk singularity
\cite{Craps:2006xq}.
  
  \bibitem{lifshitz}
  The dilation symmetry \eqn{dilation} looks like a time dependent version of the Lifshitz symmetry \cite{Kachru:2008yh}, since both  rescale some directions differently from others. However  there is an important difference. Here, the boundary metric is a standard four-dimensional spacetime and the symmetry is an isometry of this metric. In the Lifshitz case, the boundary is more complicated.
  
\bibitem{Kachru:2008yh}
  S.~Kachru, X.~Liu and M.~Mulligan,
  ``Gravity duals of Lifshitz-like fixed points,''
  Phys.\ Rev.\ D {\bf 78} (2008) 106005
  [arXiv:0808.1725 [hep-th]].
  
  \bibitem{bala}
  See also e.g. \cite{Balasubramanian:2012tu}
for a recent discussion of complex geodesics in gauge/gravity duality.

  
\bibitem{Balasubramanian:2012tu}
  V.~Balasubramanian, A.~Bernamonti, B.~Craps, V.~Ker\"anen, E.~Keski-Vakkuri, B.~M\"uller, L.~Thorlacius and J.~Vanhoof,
  ``Thermalization of the spectral function in strongly coupled two dimensional conformal field theories,''
  JHEP {\bf 1304} (2013) 069
  [arXiv:1212.6066 [hep-th]].
  
  \bibitem{sing}
This is well defined since the solutions \eqn{soln1/4}--\eqn{soln1/4b} are nonsingular at $w=0$.
  


\bibitem{Louko:2000tp}
  J.~Louko, D.~Marolf and S.~F.~Ross,
  ``On geodesic propagators and black hole holography,''
  Phys.\ Rev.\ D {\bf 62} (2000) 044041
  [hep-th/0002111].
  
  \bibitem{us}
  N Engelhardt, T. Hertog, and G. Horowitz, to appear.
 
\bibitem{ross}
  We thank S. Ross for pointing this out to us.

    
  
\bibitem{Engelhardt:2013jda}
  N.~Engelhardt and G.~T.~Horowitz,
  ``Entanglement Entropy Near Cosmological Singularities,''
  JHEP {\bf 1306} (2013) 041
  [arXiv:1303.4442 [hep-th]].
  
  
    
   \bibitem{skenderis}
 See \cite{McFadden:2009fg} for a different connection between Lorentzian AdS/CFT and inflation. 
  
\bibitem{McFadden:2009fg} 
  P.~McFadden and K.~Skenderis,
  ``Holography for Cosmology,''
  Phys.\ Rev.\ D {\bf 81}, 021301 (2010)
  [arXiv:0907.5542 [hep-th]].

\end{thebibliography}
\end{document}